# THE ANCIENT EGYPTIAN PERSONIFICATION OF THE MILKY WAY AS THE SKY-GODDESS NUT: AN ASTRONOMICAL AND CROSS-CULTURAL ANALYSIS

**Or Graur**

Institute of Cosmology and Gravitation, University of Portsmouth

Department of Astrophysics, American Museum of Natural History

E-mail: or.graur@port.ac.uk

**ABSTRACT**

The Milky Way's name and role in ancient Egyptian culture remain unclear. One suggestion is that the Milky Way may have been a celestial depiction of the sky goddess Nut. In this work, I test this association using an interdisciplinary approach. In the first part of this paper, I combine astronomical simulations of the ancient Egyptian night sky with primary Egyptian sources to map the goddess Nut onto the Milky Way. With her head and groin firmly associated by primary texts with the western and eastern horizons, respectively, I argue that the summer and winter orientations of the Milky Way could be construed as figurative markers of Nut's torso (or backbone) and her arms, respectively. In the second part of this paper, I situate Nut within the global, multicultural mythology of the Milky Way. Specifically, I show that Nut's roles in the afterlife and the autumn bird migration are consistent with similar roles attributed to the Milky Way by other cultures. Finally, I show that Nut's identification with the Milky Way may even have echoes in contemporary African conceptions of the Galaxy.

## 1 INTRODUCTION

The ancient Egyptians wove the workings of the night sky into many aspects of their culture (for a recent review, see Belmonte and Lull 2023). In Egyptian cosmology, the world, which consisted of Egypt and its immediate neighbors, was surrounded by infinite, inert waters (the abyss, Nun). The earth, personified by the god Geb, was protected from the encroaching waters by the sky, personified by Geb's sister and consort, Nut, who was held aloft by the atmosphere, represented by their father, Shu (Figure 1). The connection between the sky and the waters that it holds back is seen in the hieroglyphs that make up the names Nut and Nun. The abyss, Nun, is commonly written as either ⬡ or ⬡⬡ (*nwn* in the proto-Semitic transcription used here), while Nut usually appears as either ⬡ or ⬡ (*nwt*). Nut's name is composed of the phonetic combination ○ (*nw*) and ⌒ (*t*) along with the hieroglyph for sky, ▬ (*pt*), and the female anthropomorphic determinative 𓁐. The hieroglyph for sky also appears in the name of the abyss, and its watery nature is attested by the triple use of the hieroglyph for water, ▬ (*n*). This connection is further strengthened by the variant ⬡ used for Nut in certain versions of the *Coffin Texts* (de Buck 1935: 256d, 270f; van der Molen 2000: 204; for further discussion of this connection, see Billing 2002: 11).

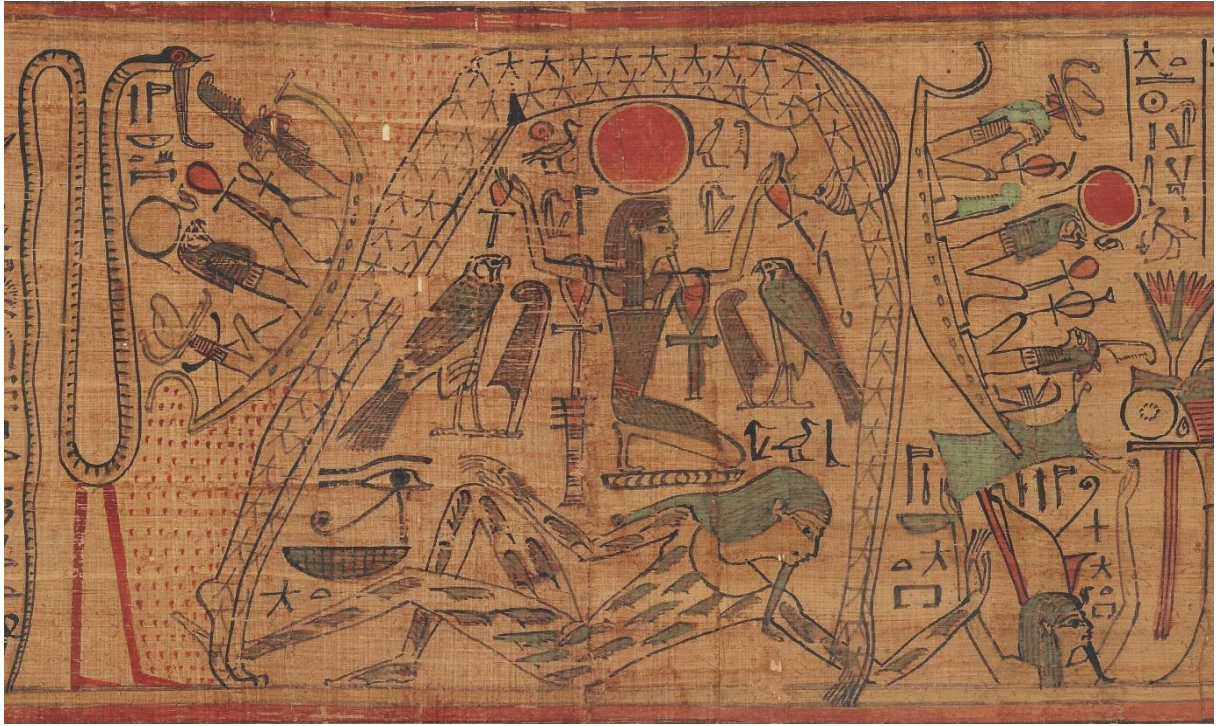

**Figure 1.** Nut, covered with stars, is arched above the earth god Geb, and held aloft by her father, the atmosphere god Shu. The rising Sun, carried by Re and colored yellow, sails up Nut's legs while the setting Sun, colored red, sails down her outstretched arms towards Osiris (courtesy: gallica.bnf.fr / Bibliothèque nationale de France, département des manuscrits, Egyptien 172).

Nut is a major figure in the Heliopolitan Ennead, i.e., the "Heliopolis Nine," a pantheon and family of gods headed by Atum, who created himself out of the abyss. Atum is the Sun, marking the central role of our star in Egyptian religion and cosmology. The Sun is also deified as the gods Re and Khepri, who are not part of the ennead; together, all three Sun gods are thought to symbolize the three daily phases of the Sun: the scarab-headed Khepri is the rising Sun, Re is the midday Sun, and Atum is the setting Sun (Allen 1988: 10). Atum created Shu (god of the atmosphere) and Tefnut (goddess of moisture). Together, Shu and Tefnut begat Geb (god of the earth) and Nut (goddess of the sky). Both siblings and partners, Geb and Nut begat Osiris, Isis, Seth, and Nephthys. While competing cosmogonies developed in other parts of Egypt (Memphis, Thebes, Hermopolis, and Esna; for a detailed discussion, see Belmonte and Lull 2023: 1-22), the Heliopolitan cosmogony, which includes Shu, Geb, and Nut, is central to this paper.

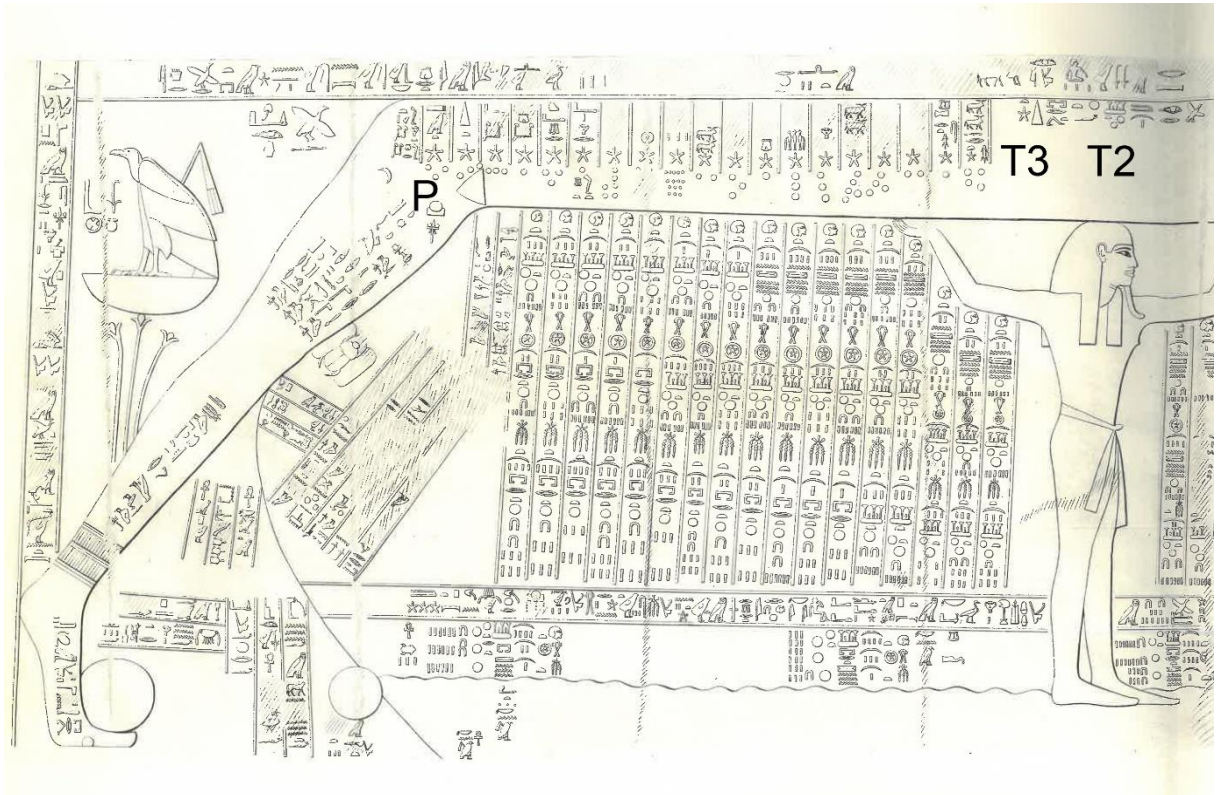

**Figure 2a.** Left half of a reconstruction of the *Fundamentals* from the Cenotaph of Seti I at Abydos (after Frankfort 1933: plate 81; courtesy of The Egypt Exploration Soceity).

Nut plays an important role in the Egyptian conception of the Solar cycle, according to which the Sun is ferried by boat across the waters of the sky from dawn to dusk. In the *Book of the Divine Cow*, it is Nut who, turned into a cow, supports Re on her back (e.g., Clagett 1992: 539-540; Hornung 1982: 41-42; Simpson 2005: 290-293). According to the *Fundamentals of the Course of the Stars* (formerly known as the *Book of Nut*, the correct name of this work was revealed by von Lieven 2007 and is hereafter referred to in short as *Fundamentals*), as the Sun sets in the west, it is swallowed by Nut, who carries the Sun/Re into the Duat (*dwȝt*; the Egyptian netherworld) inside her body. The spent Sun/Re sails through the Duat, where it encounters Osiris who makes it effective once more. Nut then gives birth to the regenerated Re and the Sun rises in the east (von Lieven 2007). Textual and visual descriptions of this process are found in monumental versions of the *Fundamentals* in the Cenotaph of Seti I in Abydos (Figure 2) and the tombs of Ramesses IV and Mutirdis, as well as ceiling paintings in the tombs of Ramesses VI and Ramesses IX, funerary papyri, and coffins.

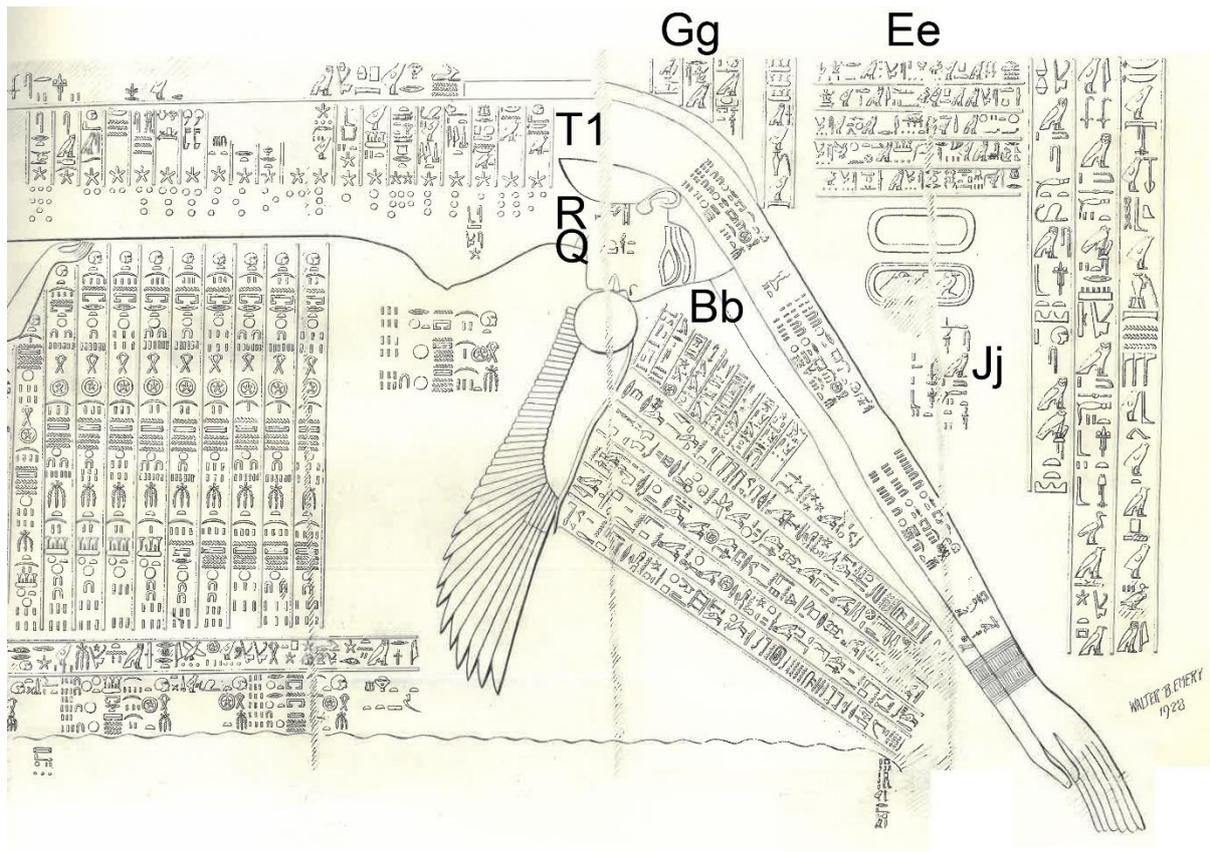

**Figure 2b.** Right half of a reconstruction of the Fundamentals from the Cenotaph of Seti I at Abydos (after Frankfort 1933: plate 81; courtesy of the Egypt Exploration Society.

## 1.1 The Egyptian Name for the Milky Way

The ancient Egyptians were highly aware of their environment, which they recorded in both textual and visual media. The latter, including sculptures, papyri, and tomb and temple decorations, are so detailed that they have been used to identify many individual species of fish and birds (e.g., Bailleul-LeSuer 2012; Brewer and Friedman 1989; Houlihan 1988). The same is true of astronomy; Egyptian texts and tomb decorations have supplied Egyptologists with clear connections between Egyptian mythology and certain astronomical objects, such as the Sun and the Moon; the planets Mercury, Venus, Mars, Jupiter, and Saturn; the star Sirius; and the constellations Orion and the Plow (however, the location of other constellations, such as those recorded in the tomb of Senenmut, is still debated; e.g., Belmonte 2003; Belmonte and Lull 2023). Due to the Egyptian's interest in the night sky, it has long been assumed that they were also aware of the Milky Way and incorporated it, in some shape or form, into their culture. Up until the modern era, when anthropogenic light pollution has obscured the Milky Way from most of humanity (e.g., Falchi et al. 2016), the opalescent band of our Galaxy was one of the most distinctive features of the night sky; most (if not all) cultures have a specific name and origin story for this band (e.g., Graur 2024; Lebeuf 1996; Krupp 1991).

And yet, no clear name, depiction, or description of the Milky Way has been found in Egyptian texts and visual media. Chatley (1941) claimed, without evidence, that the stars trailing to the right of the ram constellation on the ceiling of the tomb of Senenmut indicated the Milky Way. Based on a semantic analysis, Sethe (1936: 20) suspected the term 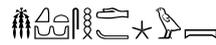 (*mskt sḥdw*) to be the Egyptian name for the Milky Way. Faulkner (1969: 71, 72n11), perhaps inspired by Sethe's use of the German name for the Milky Way, "Milchstrasse," translated this term as "Street of Stars." James Allen, in his modern translations of the *Pyramid Texts*, has suggested two competing translations: "Path of Sailing Stars" (Allen 2014: 25-26) and "Beaten Path of Stars" (Allen 2015: 51, 82, 134, 351). Conman (2003), however, argued that this term describes the ecliptic. Davis (1985) suggested a different term from the *Pyramid Texts* as the Milky Way's Egyptian name: 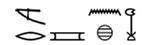 (*Mr-nḫ3*), which has been translated as "Winding Canal," "Winding Waterway," and "Shifting Waterway." Krauss (1977: 63-64), however, translated this term as "Kha Canal" and argued for its identification with the ecliptic.

Maravelia (2017a) tied *mskt sḥdw* to the Coptic name for the Milky Way, ⲙⲱⲓⲧ ⲛⲧⲉⲡⲓⲧⲟϨ. However, this term, literally "Way of Chaff" (Crum 1939: 435b; Vycichl 1983: 225; Westendorf 1977: 257) recalls a family of names for the Galaxy found across North Africa, the Middle East, the Arabian Peninsula, Persia, and Central Asia, which derive from the Arabic name "Path of Straw" (درب التبانة - *Darb al-Tabana*). Crum (1939) lists this name, in Arabic (as طريق التبن – *Tariq al-Taban*), as well as an Arabic synonym, المجرة – *al-Majara*. Thus, I find it likelier that the Coptic term is an import from Arabic rather than a throwback to *mskt sḥdw*. This conclusion has been confirmed in consultation with Alicia Maravelia and Sherin Sadeq el-Gendi (pers. comm., 2023).

Citing Neugebauer and Parker (1960: 50), Egberts (1995: 292f, note 5), and Willems (1996: 262-270), von Lieven (2007: 136-137) argued that *mskt*, which is associated with an eastern transit area of the Sun and stars, the sunset, and the gates of the Duat (based on spell 789 from the *Coffin Texts*; Faulkner 1978: 1), most plausibly refers to a celestial region that lies directly above the horizon, but whose vertical reach is as yet unknown. In conclusion, it is still unclear what, exactly, *mskt* and *mskt sḥdw* refer to.

**1.2 Identifications of Nut with the Milky Way**

Davis (1985) suggested that the Milky Way may have been associated with the Egyptian sky goddess. She noted that in the *Pyramid Texts* the sky is divided into northern and southern skies, and the Sun is mentioned crossing from one sky to the other. As Davis (1985) noted, the Milky Way can be said to divide the sky in two, and it is crossed by the Sun once a year from south to north and, six months later, from north to south. The passage of the Sun across the Milky Way mimicked, in her opinion, the myth of the sky goddess swallowing the Sun in the evening and giving birth to it once more in the morning. Davis (1985) did not name Nut outright and, confusingly, referred to the sky goddess as being "often represented in very round hippopotamus form." However, the mention of the Solar cycle myth clarifies the connection to Nut and, though Nut's earliest and most common representation is anthropomorphic, she is also depicted as a cow, a vulture, a sow, and a hippopotamus (Billing 2002: 13-24).

Kozloff (1992, 1994) also identified Nut with the Milky Way by arguing that 18th-dynasty spoons in the shape of a slim, naked young woman holding an object in her outstretched arms were depictions of Nut. In this interpretation, a spoon with a goose at its end is Nut holding Geb, who is sometimes identified with geese; when she holds a duck, it is Nut holding Geb's son; a lotus flower – the Sun. Kozloff (1992) then noted that the constellation Cygnus (the Swan) lies along the Milky Way and suggested that, since swans did not exist in ancient Egypt, Egyptian observers would have seen in the stars of Cygnus a different long-necked bird, such as a goose. With the "swimming girl" spoons in mind, Kozloff (1992) oriented Nut in the sky such that her arms stretch eastwards to grasp the constellation Cygnus.

There are several glaring problems with this argument. First, as Kozloff (1992) briefly noted, some spoons, such as those where the swimming girl holds a gazelle, do not easily fit an identification with Nut. Second, mute and whooper swans are attested in ancient Egyptian art (Houlihan 1988: 50-54), so there is no need to replace the Swan with a Goose. Third, while many cultures identify similar patterns in the night sky (such as the Pleiades star cluster or the asterisms that make up Orion and the Plough; Kemp et al. 2022), there is no evidence for an Egyptian constellation in Cygnus (Belmonte 2003). Even if there were such a constellation, there is no reason to assume it would have been that of a bird (where we see a Plow in Ursa Major, for example, the Egyptians saw the leg of a bull). Fourth, as Maravelia (2003, 2017b) noted, there are many astronomical errors throughout Kozloff (1992, 1994). Finally, and most importantly, orienting Nut so that she faces to the east is at odds with the myth of the Solar cycle and with her description in the *Fundamentals* (see Section 3, below).

Using then-available planetarium software and models of the Milky Way, Wells (1992) noted that the appearance of the Milky Way during the night of the winter solstice recalled Nut's anthropomorphic depiction. He then proceeded to locate Nut's groin in the area of the Milky Way covered by Cygnus, reasoning that Cygnus is located near a junction where the Milky Way bifurcates, and that the separate strands of the Milky Way could represent Nut's legs. Wells (1992) attempted to strengthen this argument by noting that "Cygnus, the swan, is also sometimes called the 'Northern Cross' because of its obvious appearance. Figurines of primitive societies in clay or other materials which represent the female form have been found which denote the genitalia by a cross inscribed in the appropriate position between the legs." However, as Hollis (2019: 70-91) noted, based on Ucko (1968: 192, 196, 202), this is not the case for ancient Egypt.

Wells (1992) continued to argue that, if Nut gives birth to the Sun in Cygnus close to sunrise on the winter solstice, she should swallow it when her head, in Gemini, is just above the horizon after sunset. This, according to his simulations, happens an hour after sunset on the spring equinox, nine months before the winter solstice. In his argument, this period, equal to the human gestation period, turns the myth of Re's death and rebirth from a daily event into an annual one and cements Nut's identification with the Milky Way.

The works described above provide no definitive proof that the Milky Way is a personification of Nut, and the mappings of Nut onto the Milky Way provided by Davis, Kozloff, and Wells are inconsistent with each other and, at times, with her description in primary texts. In this work, I reexamine the connection between Nut and the Milky Way. In Section 2, I describe the planetarium software I use to compare, in Section 3, the Milky Way's appearance in the ancient Egyptian night sky to Nut's descriptions in the *Fundamentals*, the *Pyramid Texts*, and the *Coffin Texts*. In Section 4, I compare Nut's roles in the formation of the Egyptian pantheon, the transition of the king to the afterlife, and the autumn bird migration with similar roles and origin stories attributed to the Milky Way by various cultures across the world. As I conclude in Section 5, I find that Nut's description in primary texts is suggestive of a figurative connection to the Milky Way, such that its summer orientation highlights Nut's torso (or backbone) and the winter orientation delineates her arms. Such a connection would place Nut comfortably within multicultural conceptions of the Milky Way.

## 2 METHODS

I use two open-source planetarium computer programs, Stellarium and Cartes du Ciel, to simulate the ancient Egyptian night sky. Stellarium (Zotti and Wolf 2023), which is used extensively for archaeo- and cultural astronomy (Zotti et al. 2021), allows users to simulate the appearance of the night sky anywhere in the world at any time between 100,000 BCE and 100,000 CE. However, its calculation of the positions of the planets is only accurate between 4000 BCE and 8000 CE, and its calculation of stellar proper motions is inaccurate before 2500 BCE (De Lorenzis and Orofino 2018). One of the distinctive features of Stellarium is its use of a digital, all-sky mosaic of the Milky Way, which provides users with an accurate image of the Galaxy as seen from Earth down to a limiting magnitude of ~14 mag (Mellinger 2009), a thousand times deeper than the human limiting magnitude of 6.5 mag. Cartes du Ciel is used to produce sky charts for amateur telescope observations, and is considered by De Lorenzis and Orofino (2018) to be more precise than Stellarium.

In Cartes du Ciel, I have chosen to show the stars as they would look to a naked-eye observer. Note that the Milky Way in Cartes du Ciel is only a line model of the Galaxy that does not represent its overall brightness or the variations in brightness along its full extent. In Stellarium, I have chosen to display the night sky with Bortle class 2 light pollution, which indicates a typical dark-sky location. Regrettably, the Milky Way, which is easily identifiable in the Stellarium application, is barely noticeable in screen captures. To remedy this issue, I have artificially increased the brightness and contrast of the images produced by Stellarium. As a result, they should not be construed as authentic views of the Milky Way. Instead, they should only be contrasted with each other to highlight the changes in orientation, relative brightness, and nearby constellations between different seasons.

In the following section, I compare the appearance of the Milky Way in the night sky with descriptions of Nut in the *Pyramid Texts*, the *Coffin Texts*, and the *Fundamentals*. Here, I rely on Allen (2015) for a modern translation of the *Pyramid Texts* and on Faulkner (1973, 1977, 1978) for translations of the *Coffin Texts*. Spells from these sources are listed as PT and CT, respectively. For both sources, I follow the convention of substituting NN for the king's name. For the *Fundamentals*, I rely on the comprehensive translation of von Lieven (2007). Unless noted otherwise, translations from the German text of von Lieven (2007) are my own.

With these sources in mind, I have simulated the Egyptian night sky as viewed from both the Valley of the Kings (N 25°44′23.30″, E 32°36′06.81″, 190 m elevation) and Giza (N 30°00′33.98″, E 31°12′31.0″, 19 m elevation) during three periods: 1279 BCE (the final year of Seti I's reign, corresponding to the age of the first witness of the *Fundamentals*), 1880 BCE (the period to which von Lieven 2007 attributes the assembly of the *Fundamentals*), and 2300

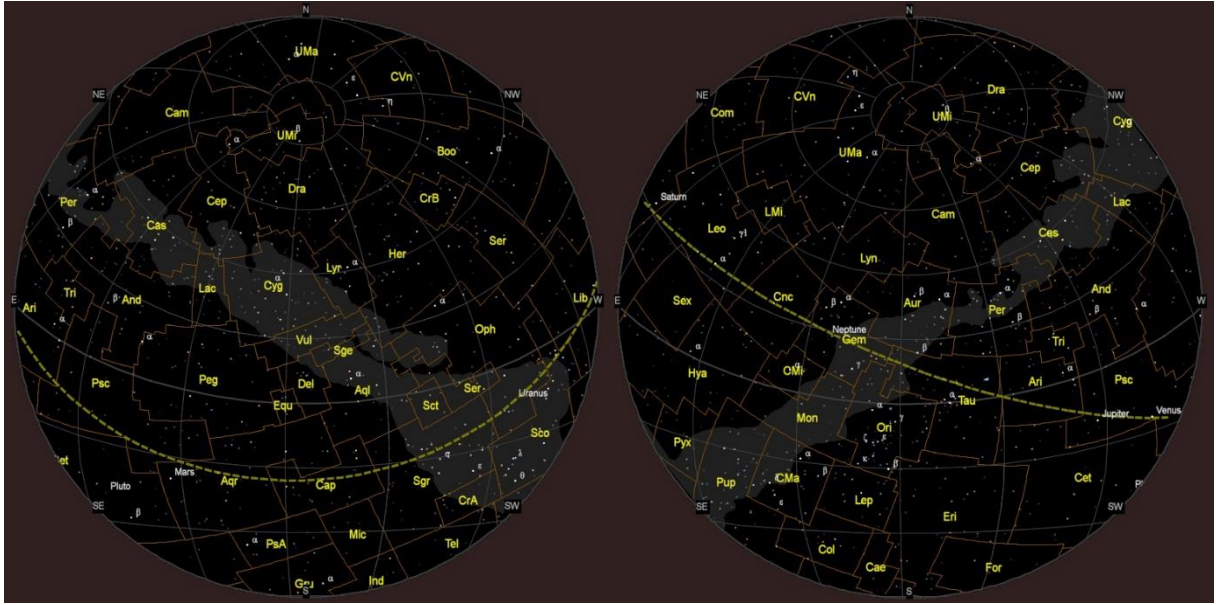

**Figure 3.** Cartes du Ciel sky charts of the Egyptian night sky in Giza on 1 July 1880 BCE, 00:00 (left) and 1 January 1880 BCE, 20:00 (right). Note the change in orientation of the Milky Way (the gauzy diagonal band).

BCE (the period of the Old Kingdom to which the oldest witnesses of the *Pyramid Texts* are attributed).

## 3 ASTRONOMICAL ANALYSIS

The appearance of the Milky Way in the Egyptian night sky changes with the seasons and throughout the night (Figures 3-4). If we consider the Milky Way only when it is highest in the sky on any given night (as that is when it would be brightest and most prominent), we find that, in general, during the summer months (roughly April to August, during 1880 BCE), the Milky Way stretches from the northeast to the southwest. During this period, the brightest swath of the Milky Way is visible, as an observer would be looking inwards towards the center of the Galaxy. During September, the Milky Way can be seen to flip its orientation throughout the night. After the Sun has set, the Milky Way stretches from the northeast to the southwest, but as the night wears on the Milky Way swings across the sky, briefly assuming an east-west orientation before finishing the night stretching from the southeast to the northwest. The latter orientation remains dominant throughout the winter months (roughly October to February, 1880 BCE). During this period, an observer would be looking outwards towards the edge of the Galactic disk, and hence the Milky Way is at its faintest. Throughout the nights of March, the Milky Way starts out with its winter orientation, then dips close to the horizon before rising again with its summer orientation. This trend is visible in all three eras tested here.

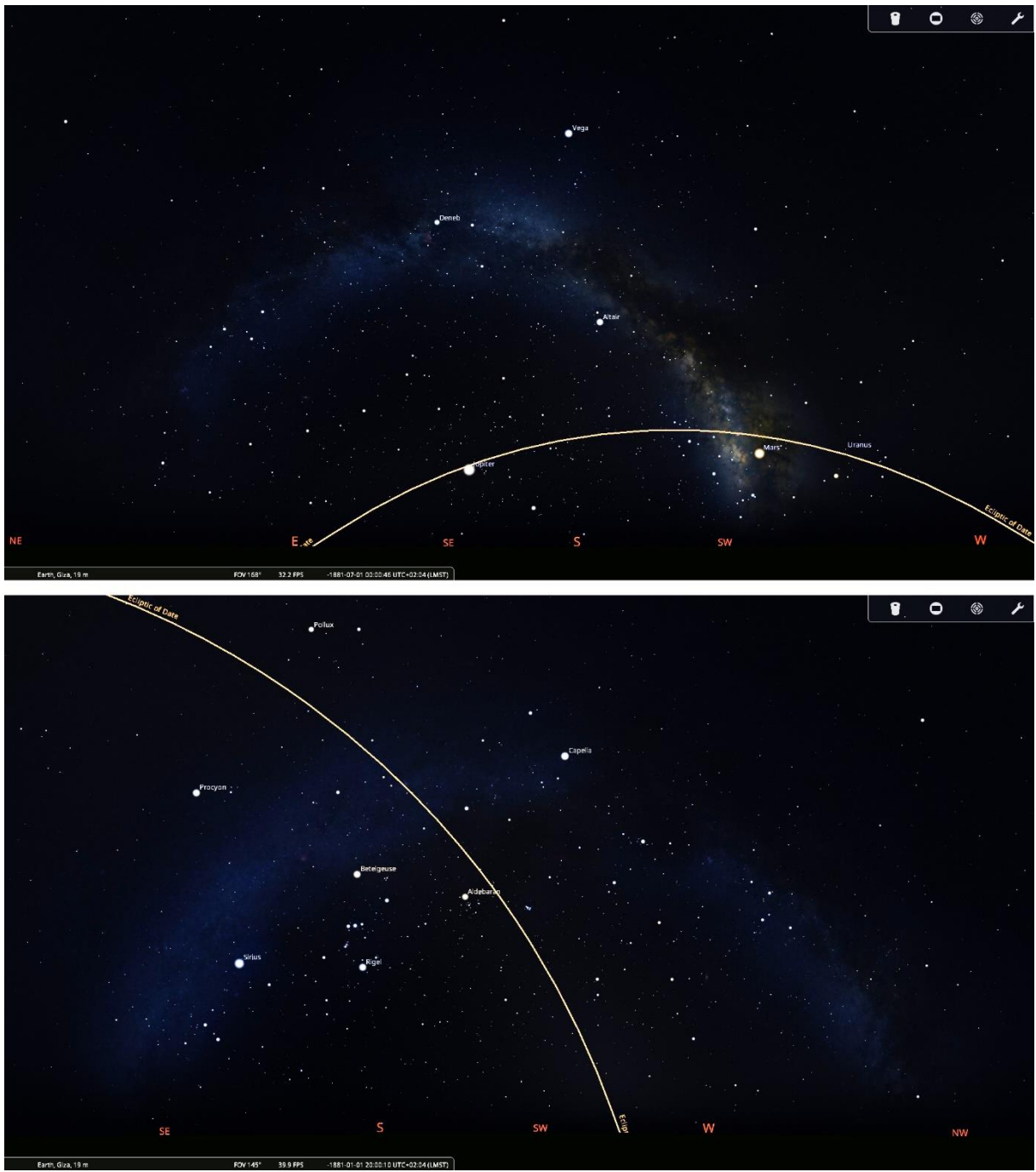

**Figure 4.** Stellarium views of the Egyptian night from Giza on 1 July 1880 BCE, 00:00 (top) and 1 January 1880 BCE, 20:00 (bottom). The overall brightness and contrast of the images has been increased by the author. Note the change in the orientation of the Milky Way between the two dates, consistent with Figure 3. Note also the difference in brightness between the two appearances of the Milky Way and the locations of Sirius and Orion.

The orientation of the Milky Way in the night sky is critical, as it must be compared to the physical description of Nut provided by the *Fundamentals*. Following the mapping of the text of the *Fundamentals* from the Cenotaph of Seti I (Figure 2; for the complete mapping, see Neugebauer and Parker 1960: figure 20), von Lieven (2007) translates text Gg, next to Nut's head as: "Her head is the western horizon, her mouth is the west." Text Q, above Nut's mouth, reinforces the latter by reading "Western horizon," while text R, above, reads "Evening." Text Jj, next to her arms, reads: "Her right arm is on the northwest side, [her left arm] is on the southeast side." And text P, next to Nut's groin, reads: "Eastern horizon." Nut's general orientation is repeated at the end of text Ee: "Her rear is in the east, and her head is in the west."

That Nut's head lies in the west and her rear in the east is only to be expected from the myth of the Solar cycle. The orientation of her arms, however, is very specific. Unlike the two-dimensional visual representations of Nut in the *Fundamentals*, on coffins, and in funerary papyri, the text of the *Fundamentals* paints a three-dimensional image where Nut's arms stretch to her sides at 45-degree angles. Moreover, the orientation of the arms is either parallel or perpendicular to the Milky Way (whether observing it during the winter or summer months, respectively). This is the most suggestive piece of evidence for a connection between Nut and the Galaxy. At the same time, the changing orientation of the Milky Way throughout the year presents a serious dilemma. During the summer months, the orientation of Nut's arms would imply that the Milky Way represents Nut's body, or at least some part of it. But during the winter months, the flipped orientation of the Milky Way would be consistent with representing only her arms. In the following subsections, I analyze the evidence for and against associating Nut with either (or both) of these orientations of the Milky Way.

## 3.1 Summer Orientation: Nut's Torso

As von Lieven (2010) notes, the list of decanal stars (specific stars and asterisms used as a star clock; see Belmonte and Lull 2023: 123-132) along Nut's torso in the *Fundamentals* (texts T1, T3) is linked to one specific date. This date, found in text T2 in the center of Nut's torso, was translated by Neugebauer and Parker (1960: 54) as: "All these stars begin in the sky in I Akhet when Sothis rises" and by von Lieven (2007: 61) as: "What is done at I Akhet in view of the emergence of Sothis." The month of I Akhet marks the beginning of the inundation of the Nile, which occurs in July. This is also the month that sees the heliacal rising of Sirius, known in Egypt as Sothis or Sopdet. Thus, it is possible that the description of Nut in the *Fundamentals* is also linked to that specific moment in time.

Further evidence for the importance of the summer season comes from the *Pyramid Texts*, in which Nut gives birth to the rejuvenated king in the same area where Orion, Sirius, and the morning star are seen to rise. For example:

> The sky conceives you with Orion, the morning-star gives you birth with Orion. Live! Live, as the gods have commanded you to live.
>
> You ever go up with Orion in the eastern arm of the sky, you ever go down with Orion in the western arm of the sky. Sothis, whose places are clean, is the third of you two: she is the one who leads you two from the Field of Reeds to the perfect paths in the sky.
>
> PT 442 (Allen 2015: 112)

> Ho, NN! You are the big star that is Orion's companion, who travels the sky with Orion and rows the Duat with Osiris. You emerge in the eastern side of the sky, renewed at your proper season and rejuvenated in your time, Nut having given you birth with Orion, the year having put your headband on you with Osiris.
>
> PT 466 (Allen 2015: 128)

> When the sky becomes pregnant with wine and Nut has delivered her daughter the morning star, I raise myself, the third (companion) of Sothis of clean places, having become clean in the Duat lakes and purged in the jackal lakes.
>
> PT 504 (Allen 2015: 161)

Thus, it is conceivable that, after being swallowed by Nut in the west, the Sun would travel through Nut's torso – along the Milky Way – until it reached the eastern horizon, where during the summer months, it was reborn along with Orion and Sirius. One could even imagine the Sun hidden by the Great Rift, the dark band of interstellar dust that bisects the Milky Way during the summer, as it journeyed through the Duat. The latter, however, is pure conjecture.

### 3.2 Winter Orientation: Nut's Arms

The Milky Way's winter orientation is precisely the orientation described by the *Fundamentals* for Nut's arms. Several PT and CT spells mention Nut's arms outright in connection with her roles of enveloping and protecting the king on the one hand and—imperative for this discussion—her role of helping the deceased king up to the sky. For example:

> Nut, of long hair and pendulous breasts, has given her arms toward you. She continually shoulders you to the sky and cannot drop you to earth. She continually gives birth to you, NN, like Orion, and makes your abode at the fore of the Dual Shrines.

PT 697 (Allen 2015: 291)

A ladder to the sky shall be put together for you and Nut will extend her hands towards you, you shall navigate on the Winding Waterway and sail in the eight-boat.

CT 62 (Faulkner 1973: 58)

O Nut, spread yourself over me when you enfold me with the life which belongs to you; may you fold your arms over this seat of mine, for I am a languid Great One. Open to me, for I am Osiris; do not close your doors against me, so that I may cross the firmament and be joined to the dawn, and that I may expel what Re detests from his bark.

CT 644 (Faulkner 1977: 220)

In the winter, Sirius (Sothis/Sopdet) and Orion are high in the sky. The segment of the Milky Way that is directly adjacent to these constellations, though dimmer than the segments observable in the summer, may still be bright enough to see clearly during the winter. Such an interpretation may be consistent with the following CT spell:

May NN be encircled by Orion, by Sothis and by the Morning Star, may they set you within the arms of your mother Nut, may they save you from the rage of the dead who go head-downwards…

CT 44 (Faulkner 1973: 36)

Contrast the last spell with PT 442, quoted in the previous section. On the one hand, in the first part of the spell the king is described as being conceived together with Orion, which is consistent with the summer rising of this constellation shortly before dawn (and with the mention of the Morning Star in CT 44). On the other hand, the second part of the same spell describes the king as going up with Orion in the "eastern arm of the sky" and going down with Orion in the "western arm of the sky." Orion's setting in the west is only visible during the winter months.

While the imagery of Nut extending her arms towards the king is consistent with the winter appearance of the Milky Way, there is always the possibility that the imagery of the PT and CT spells is not literal but simply figurative.

### 3.3 A Dynamic Milky Way vs. a Static Nut

Kozloff (1992, 1994) and Wells (1995) assumed that the Galaxy represented Nut's body and attempted to locate her head and groin along the Milky Way. I posit that such attempts are inconsistent with the conception of Nut as the sky, the myth of the Solar cycle, and the textual description of Nut in the *Fundamentals*. Any attempt to tie the Milky Way to Nut must preserve the following properties of the sky goddess:

1. Nut represents the entire sky at any given moment, during both day and night. While visual representations of Nut as the day sky and the night sky exist (e.g., in the *Book of the Day* and the *Book of the Night*), they are still representations of the same goddess, not two separate entities.
2. Nut swallows the Sun as it sets and gives birth to it as it rises. This happens every day, so Nut must always be oriented the same way in the sky, with her head to the west and her rear to the east.
3. Throughout the night, Nut also gives birth to the decanal stars as they rise in the east and swallows them again as they set in the west.

The last point is made clear in two extracts from the *Fundamentals*. In text Bb from the chapter devoted to Nut and the Solar cycle, we find the following description (von Lieven 2007: 72):

> The majesty of this god enters through her mouth into the interior of the Duat. After that, he goes forth and travels inside her. These stars enter behind him and emerge behind him. They rush to their places.

On the matter of the stars that enter behind the god (Re) and emerge behind him, the writer of the pCalrsberg I witness of the *Fundamentals* comments (von Lieven 2007: 73): "With him these stars set and with him they rise, that is, at their time of setting, because that is the time when they cannot be seen. They usually rise and set with him."

Similarly, von Lieven (2007: 80-81) notes that, like the Sun, the decanal stars are also treated as Nut's children, and their temporary invisibility is explained by her swallowing them, garnering her the epithet "the sow eating her piglets." This process, required to regenerate the stars, is difficult for Nut's brother and consort, Geb, to accept. Of this, the text says: "Let him not contend with her because she eats (her) children," and the pCarlsberg I commentator adds: "He has no contendings with her, … because she causes her children, that is the stars, to set" (translations by von Lieven 2010: 143). These descriptions require that Nut give birth and devour the decanal stars that rise and set throughout the night, which explains why the *Fundamentals* uses the eastern and western *horizons* instead of the cardinal points to describe her head and groin.

Together, Nut's description in the *Fundamentals* and her roles in Egyptian cosmology require that she remain *static* throughout the day and night. The sky is always there (as are the earth—Geb—and the atmosphere—Shu), so Nut must always be there as well. Her groin must always be in the east so that she can give birth to the decanal stars that rise throughout the night and to the Sun that rises every morning, and her mouth must always be in the west, ready to swallow the Sun and the stars as they set throughout the night. Hence, Nut's head and groin *cannot* be located along the Milky Way, whose appearance changes throughout the night as different segments of it rise and set.

Instead, the Milky Way could be thought of as a physical manifestation of Nut's torso or backbone during the summer months and of her arms during the winter months. For this interpretation to work, we must account for the tension between Nut's static nature and the dynamic appearance of the Milky Way throughout the night. This tension may be resolved, to a certain extent, by noting that, although different segments of the Milky Way rise and set throughout the night, overall, the Milky Way remains a continuous phenomenon in the night sky, one that, from dusk to dawn, retains the appearance of an arch connecting one horizon to the other.

The latter solution requires the ancient Egyptians to have viewed the Milky Way as a continuous constellation, all of which was identified with Nut. Though I have no direct evidence to support such an assumption, there are precedents from other cultures for identifying the entirety of the Milky Way as a single, monolithic constellation. The ancient Greek name for the Galaxy, *kuklos galaxías* (κύκλος γαλαξίας), literally means "milky circle," and recalls the shape of the entire Galaxy on the sky. On the other side of the world, the Navajo (*Diné*) consider the entire Milky Way (*Yikáísdáhá*) to be a single constellation (Maryboy and Begay 2010: 55), which they mark as a zig-zag pattern in dry paintings. Thus, while I cannot say whether it is *plausible* that the ancient Egyptians identified Nut with the entire Milky Way, it is at least *possible*.

A simpler solution to this problem is to abandon any attempts to explain the Milky Way as a *literal* manifestation of Nut in the sky and instead view it as a *figurative* highlighting of one of her aspects. This solution stems from the figurative associations of many Egyptian gods with various animals and natural phenomena. As noted in Section 1.2, though Nut is most commonly represented as a woman, she is also depicted as a cow, a sow, a hippopotamus, and a vulture. The cow (also associated with the goddess Hathor) and hippopotamus (more commonly linked to the goddesses Taweret and Ipet) highlight Nut's maternal aspects. The vulture, which as a hieroglyph is used in the construction of the most common word for "mother," may also signify Nut's protection of the deceased king (Billing 2002: 17). Her likening to a sow, on the other hand, is a figurative explanation for why she swallows the stars to which she earlier gave birth. In this context, the Milky Way is a figurative marking of Nut's role as the sky. Instead of *being* Nut's body, it only *highlights* it. This interpretation allows Nut to remain static and ever-present, while the changing views of the Milky Way highlight different parts of her throughout the year. In the winter, the Milky Way delineates Nut's arms, while during the summer months, it sketches out her torso (or backbone).

**4 CROSS-CULTURAL ANALYSIS**

A survey of the names and origin stories given to our Galaxy by diverse cultures around the world reveals a small number of recurring motifs. The Galaxy is often seen as a river, a road, a god, an animal, or a conduit between this life and the next (or as some combination of all five motifs). One must be careful when comparing one culture to another, especially when those cultures are separated by oceans and centuries. However, the similarities in depictions of the Milky Way between different cultures around the world are striking and, in my opinion, reveal an underlying shared human conceptualization of the night sky, dressed by the trappings of individual cultures.

The Egyptians, for examples, may not have been the only culture to visualize the Milky Way as a sky goddess who gives birth to other gods and, especially, other celestial objects. Several of the pages in the Codex Borgia, a pre-Colombian pictorial manuscript composed by the Tlaxacaltec people of the Puebla-Tlaxcala Valley in Mexico, include images of elongated beings covered with stars. These beings have been identified as the Milky Way as well as the goddess Citlalicue (Star Skirt), the "goddess of the stars," and the mother of several gods, including the Venus god Quetzalcoatl (Milbrath 1988; 2013: 80n40). In Figure 5, Quetzalcoatl is shown cutting his way out of the belly of Citlalicue; this and other, similar scenes in Codex Borgia 29-46 are interpreted as Venus crossing the Milky Way.

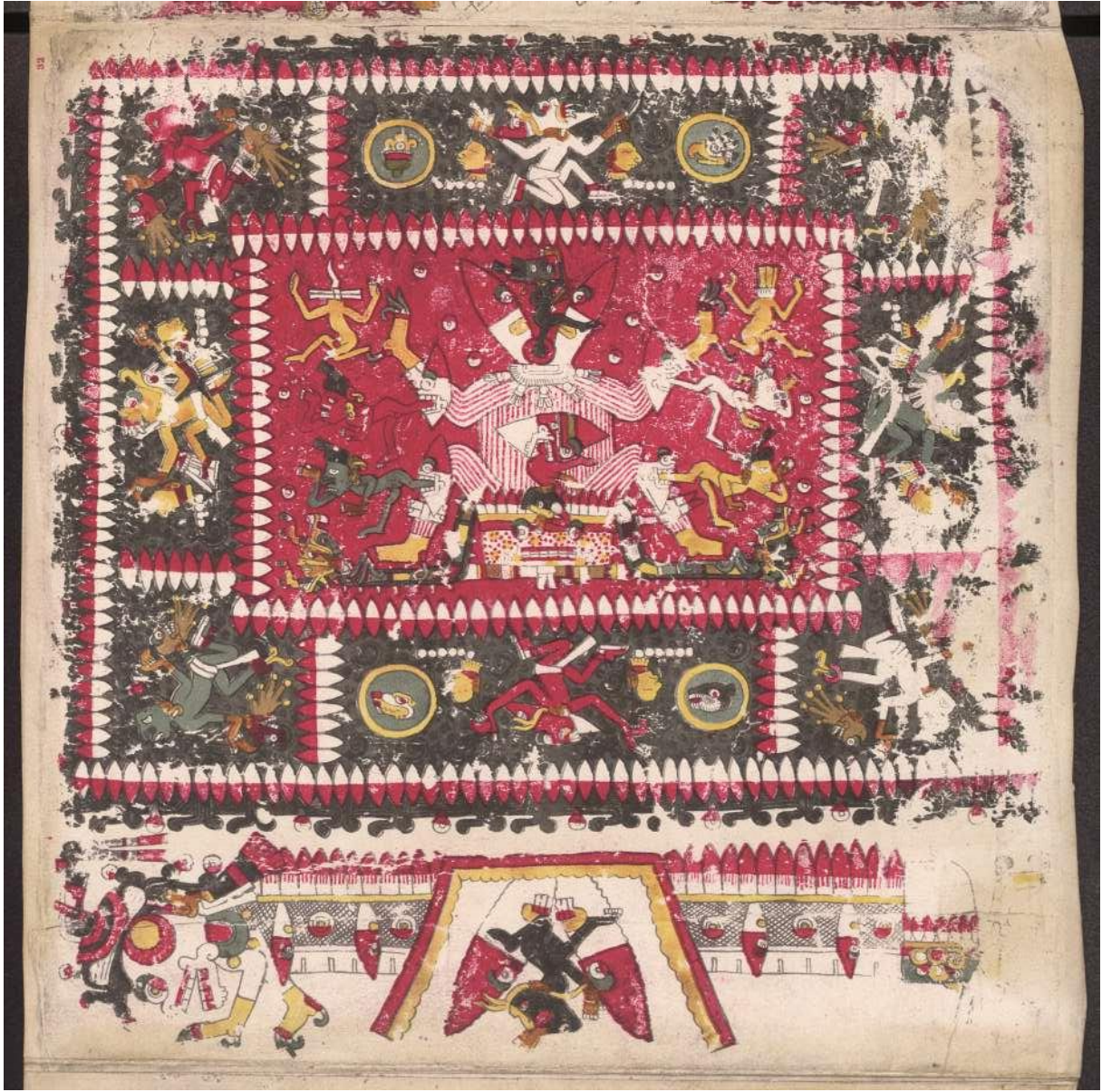

**Figure 5.** The lower border of p. 32 of the Codex Borgia is taken up by the star-studded goddess Citlalicue giving birth to Quetzalcoatl. This imagery is interpreted as Venus crossing the Milky Way (courtesy: Ancient Americas at LACMA (ancientamericas.org) in conjunction with Univeritätsbibliothek Rostock, Bibliothek der Berlin-Brandenburgischen Akademie der Wissenschaften (BBAW) and Staats- und Universitätsbibliothek Hamburg, with thanks to Michael Dürr FAMSI project coordinator, Mr. Rosenau of Mikro-Univers, Ms. Danielewski and Dr. Thiemer-Sachse of the Free University of Berlin).

There is no reason to assume that the creators of the Codex Borgia were ever in contact with the ancient Egyptians, so it is intriguing to see similar imagery—an elongated, star-studded goddess—and similar mythologies—the Milky Way giving birth to a major deified celestial object—surface in two unconnected continents thousands of years apart. In the sections below, I show that Nut's roles in the afterlife and the autumn bird migration are consistent with similar depictions of the Milky Way in other cultures across the world, and that echoes of Nut's identification with the Milky Way may still be found in Africa today.

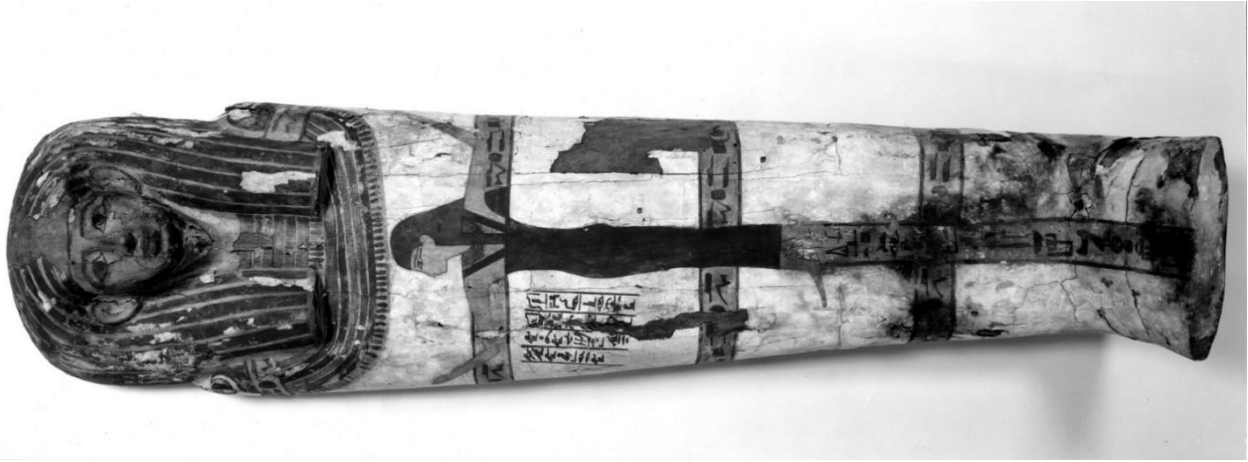

**Figure 6.** Coffin of Ahhotep Tanodjmu (ca. 1550-1458 BCE). With outstretched arms, Nut protects the deceased (courtesy: The Metropolitan Museum of Art, New York; Rogers Fund, 1912).

### 4.1 Nut and the Milky Way as a Conduit to the Afterlife

Nut plays several important roles in the transition of the deceased king to the afterlife, which echo many of the roles attributed to the Milky Way by other cultures. These duties, laid out in detail by Hollis (2019), include protecting the deceased king and reconstituting his body, giving birth to the rejuvenated king, and caring for him as a mother cares for her baby (see, e.g., PT 446, 447, 450, and 594). Nut's protection of the deceased is given physical form through her identification with the sarcophagus, coffin, and tomb (Allen 1988: 17; though note that Allen 2015: 84 translates the same words as "entombment," "tomb chamber," and "tomb's superstructure"), and depictions of Nut are commonly found inside all three of these structures (Figure 6). However, Nut's most important role, in this discussion, is the assistance she provides the king along his journey to the sky, where he eventually joins the "imperishable stars" (certain northern-hemisphere stars that are always visible in the Egyptian sky; see Belmonte and Lull 2023: 219). The *Pyramid Texts* provide several examples of this role:

> The two skies have gone to me, the two lands have come to me.
> I have stepped on the green vine under Geb's feet, and I trample Nut's paths.
>
> PT 332 (Allen 2015: 73)

> I go thereby unto my mother Nut and ascend on her in her identity of the ladder.

PT 474 (Allen 2015: 133)

I have gone up on Shu, I have climbed on the wing of Evolver. Nut is the one who has received my arm, Nut is the one who made a path for me.

PT 624 (Allen 2015: 241)

Nut's role as a ladder is also attested in CT 62 (Section 3.2). Her role as a path is also mentioned in CT 945 (Faulkner 1978: 84), which connects between every organ of the deceased and a specific divinity, e.g., "… My nails are Geb, … My thighs are Shu and Tefnut, … My path is Nut." A ladder or path for the deceased to reach the sky echoes conceptions of the Milky Way as a conduit between this world and the next found the world over (e.g., Krupp 1995; Lebeuf 1996). The following are merely a few examples.

Many Native American peoples across North America view the Milky Way as a road along which the spirits of the dead travel to the afterlife. The souls (*tasoom*) of the Cheyenne are said to travel toward the home of *Hemmawihio* (The Wise One Above), an all-knowing high god represented by the Sun, via the Milky Way, which is known as *ekutsihimmiyo* (Adamson Hoebel 1960: 86-87). The Lakota name for the Milky Way is *Wanáǧi Thacháŋku*, the Spirits' Road, which the Lakota follow to heaven when they die (Hollabaugh 2017: 70-72). The Pawnee come to this life as the children of stars and, when they die, become stars once more. The stars of the Milky Way are the ancestors of the Pawnee moving from this world to the next (Pawnee Nation Tribal Historic Preservation Office, pers. comm., 2022).

Similar views of the Milky Way are found across Mesoamerica (Milbrath 1999: 41). The dead of the Yucatec Maya travel along the Milky Way at night (Sosa 1985: 432). The Quiché Maya see the Galaxy as two of four cosmic roads. Of these, the Black Road (*Q'eqa b'e*) or Road of Xibalba (*Ri b'e xib'alb'a*), which is identified with the dark band of the Milky Way's Great Rift, leads to the underworld (Tedlock 1985: 36, 337, 354). Similarly, the Lacandón call the Milky Way the "white way of our true lord," *Hachäkyum*, the ruler of heaven populated by the dead (Duby and Blom 1975; Rätsch 1985: 37).

A final example demonstrates how the same culture can invest the Milky Way with overlapping roles. In China, the Milky Way is known as the Celestial River (天河 - *tianhe*) or Silver River (銀河 - *yinghe*). In its origin story, the Celestial River is drawn across the sky by the Emperor of Heaven to separate the Cowherd (牛郎 – *Niulang*) and the Weaver Girl (織女 - *Zhinü*), star-crossed lovers symbolized in the night sky by the stars Altair and Vega, respectively, which lie on either side of the Milky Way. At the same time, Huang (2012: 265-266) notes that the Celestial River is also "a recurring theme in Daoist bathing rites for the souls, as reflected in hymns calling upon the Heavenly Worthy to pour water from the Celestial River to bathe the summoned souls." In other words, the Celestial River acts as both a river and as a conduit for the souls of the dead.

## 4.2 Nut and the Autumn Bird Migration

Text Ee of the *Fundamentals* contains a fascinating description of what can only be described as the autumn bird migration into Egypt (von Lieven 2007: 76-78):

> These birds are like this: their faces are human faces, (while) their shape is that of birds. (Each) one of them speaks to his comrade in human language. After they have come to eat plants and to feed <in the> marshes in Egypt, they settle down under the light of the sky. Then they change into their bird form. The nesting place(?) in *ḳbḥw*. The primeval darkness, the *ḳbḥw* of the gods, the place where the birds come from. This is on her [Nut's] northwest side to her northeast side. It is open towards the Duat, located on her northern side.

As von Lieven (2007: 156-157) notes, the description of the birds in this passage is consistent with that of the *ba*, an important aspect of a person that

> … encompassed the powers of that entity. It was the vehicle by which they were manifested as individuals … (Taylor 2001: 20).

Like other denizens of the Duat, the *ba*, depicted in visual media as a human-headed bird, were free to leave the netherworld and visit the earthly realm as they pleased. To the Egyptians, who had no knowledge of lands to the north of the Mediterranean Sea, the migrating birds would have seemingly arrived out of the endless waters that surrounded the world (Allen 1988: 6). Here, they emerge from the Duat in *ba* form and morph into birds once they reach Egypt.

The autumn migration, which begins in August and stretches to December, sees millions of birds cross the Mediterranean or fly along the Jordan valley before entering Egypt and proceeding south along the Nile valley and the coast of the Red Sea. To Egyptians, the birds would seem to arrive from the north, from the northwest to the northeast, as described in the text above. The interval of the migration coincides with the period of the year when the orientation of the Milky Way flips. At the start of the migration, the birds would seem to follow the Milky Way from the northeast down to the southwest, while during the end of the migration season, the birds arriving from the northwest would follow the Milky Way towards its southeastern edge.

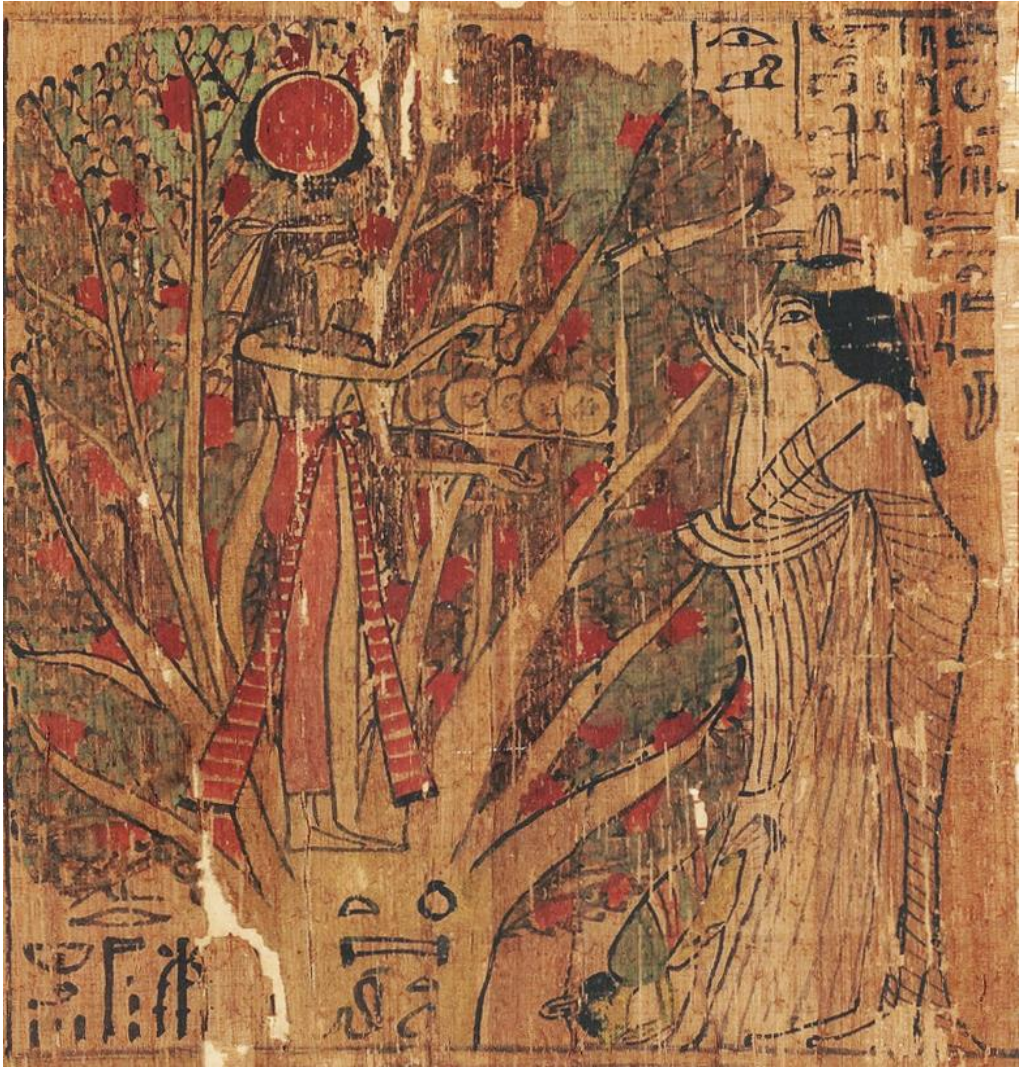

**Figure 7.** Nut (red), perched in a sycamore tree, provides food and drink to the deceased chanteuse Tentamon (white) and her *ba*, the human-headed bird next to the deceased. (courtesy: gallica.bnf.fr / Bibliothèque nationale de France, département des manuscrits, Egyptien 172).

Several funerary papyri depict Nut, perched in a sycamore tree, offering food and drink to the deceased and their *ba* (Figure 7). Furthermore, the Duat, from which the birds emerge in *ba* form, is described as both an underworld and as the inside of Nut's body. The latter is most evident in descriptions of the Sun's passage through the Duat after being swallowed by Nut. With this interpretation in mind, the birds emerging from the Duat in *ba* form could be said to be emerging from Nut. The passage, however, notes that the entrance to the Duat is "on her northern side," not the west, where the Sun sets daily. As von Lieven (2007: 156-157) suggests, this tension may be due to evolving perceptions of the Duat between the era of the Old Kingdom and the era(s) when the *Fundamentals* and its commentaries were written.

With links between Nut, *ba* birds, and the autumn bird migration already established, a link between the bird migration and the Milky Way would strengthen the link between Nut and the Galaxy. Such a link would be reminiscent of conceptions of the Milky Way found in several cultures in northeastern Europe, who regard the Milky Way as the path along which birds migrate before winter (Krupp 1995). This link still exists today in the name given to the Galaxy by Finland, Estonia, and several Baltic states: "Birds' Path" (e.g., *Linnunrata* in Finnish or *Paukščių Takas* in Lithuanian).

### 4.3 The Milky Way in Africa

Given the millennia-long existence of ancient Egypt and its influence on its neighbors, it is intriguing to check whether any echoes of Nut can be identified in contemporary conceptions of the Milky Way throughout Africa. An echo of Nut may exist, for example, in Hebrew. The final chapter of *Genesis* begins with a description of Jacob's death and burial (*Genesis* 50:1-3, New International Version; see also *Genesis* 50:26):

> Joseph threw himself on his father and wept over him and kissed him. Then Joseph directed the physicians in his service to embalm his father Israel. So the physicians embalmed him, taking a full forty days, for that was the time required for embalming. And the Egyptians mourned for him seventy days.

The Egyptian mummification and mourning rituals are recorded to have lasted 70 days, 40 of which were dedicated to drying the corpse before it was wrapped and placed in a coffin (the length of the mummification ritual is attested, for example, by the pCarlsberg I witness of the *Fundamentals* where, to explain why the decanal stars spend 70 days in the Duat after they set, the commentator notes that: "He loses his evil for 70 days, that is, after 70 days it is said: 'Therefore they are placed in the embalming workshop for 70 days.'"; von Lieven 2007: 87). This has led some scholars to connect between the embalming of Jacob and the Egyptian mummification ritual (e.g., Galpaz-Feller 2006). I note that the word "embalmed" in this verse is a translation of the Hebrew root חנט (*ḥnt*), and those who are embalmed are described as חֲנֻטִים (*ḥȝnwtîm*). Even in modern Hebrew, חָנוּט (*ḥȝnwt*) means "mummy" in the Egyptian sense of the word. I cannot help but wonder whether this Hebrew term is an echo of Nut (*nwt*), derived from her identification with the coffin and central role in Egyptian burial rites.

As one of the largest continents in the world, Africa is home to many peoples, cultures, and languages. Thus, it is no surprise to find a wide breadth of names and mythologies for the Milky Way (e.g., Alcock 2014; Langercrantz 1952; Warner 1996). Though there is no clear identification of Nut in any of these names, some of them are reminiscent of the Egyptian conception of Nut and her cosmological roles.

The Nyae Nyae !Kung, the Tshimbaranda !Kung, the Nharo, and the !Ko call the Milky Way "Backbone of the Night" (Lee 1984: 19; Marshall 1999: 256-257). A similar term is found among Xhosa speakers, for whom the Milky Way is *umnyele wezulu*, lit. "Bristles of Heaven" (Fischer 1985: 381), a name that recalls "raised bristles along the back, as on an angry dog" (Soga 1931: 419). This conception would be consistent with an identification of the Milky Way as Nut's back or backbone and with part of CT spell 660:

> O knife which is on Nut, O flatterer(?) of faeces, do not use your hand against me, do not kiss the backbone(?) of the souls at the front of the sky, because they have indeed flown up to the sky as falcons, and I am on their wings…
>
> CT 660 (Faulkner 1977: 231)

However, envisioning the Milky Way as a backbone may be a general archetype of the Milky Way, similar to the road and river archetypes. Evidence for this comes from southwestern North America, where, according to Harrington (1908: 41, 51) the people of Tewa, Taos, and Jemez Pueblos also call the Milky Way "Backbone of the Universe" (*'Opatuk'u*).

Stronger echoes are found elsewhere. According to a particular Tswana informer, the Milky Way, called *molalatladi*, supports the sky, preventing it from falling down to Earth, and controls day and night (Alcock 2014: 255). Perhaps the strongest echo comes from the G/wi of Botswana. Silberbauer (1981: 108) relates the following explanation for what happens when the Sun sets:

> When the sun sinks in the west, darkness is caused by the shadows of the trees of that far country. "N!adima makes the sun come down to that country so that it may cause these shadows and there may be darkness for man and animal to sleep and for the nocturnal animals to move about in. N!adima does this also so that he may catch the sun and eat its top [or outer] body." It is not clear whether he does this to derive nourishment or whether he does it to prevent the sun from getting up again in the night. Both reasons were advanced separately by different informants and possibly both purposes are served by this act. The "true body" of the sun is then carried by N!adima to the far country in the east … and it rises up into the sky the next morning, having regrown its "top body." The Milky Way is one of the paths along which N!adima carries the sun.

N!adima's tale mirrors the Egyptian description of the Solar cycle: swallowed by Nut (N!adima) as it sets in the west, being carried through the Duat inside Nut (along the Milky Way) where it is re-enervated by Osiris (re-grows its "top body"), to once again rise in the east. I know of no connection between the G/wi, who live in the southern part of Africa, and the ancient Egyptians. Moreover, as I have noted before, similar conceptions of the Milky Way have developed independently in far more distant lands. All I can say is that the similarities are intriguing and deserve a closer look.

## 5 CONCLUDING REMARKS

I have presented two arguments for associating the Milky Way with the ancient Egyptian sky goddess Nut. In the first part of this paper, I show that the description of Nut in the *Fundamentals* (formerly known as the *Book of Nut*) is consistent with the appearance of the Milky Way in the ancient Egyptian night sky. I argue that Nut's cosmological roles require her to be ever-present and static in the night sky. Since her head and groin are associated with the western and eastern horizons, respectively, these organs cannot be located along the Milky Way. Instead, I suggest that the Milky Way acts as a figurative mark of Nut's role as the sky and highlights different parts of her throughout the year. By comparing simulations of the night sky with primary texts from the *Fundamentals*, the *Pyramid Texts*, and the *Coffin Texts*, I suggest that the winter orientation of the Milky Way traces Nut's arms, while the summer orientation outlines either her torso or her backbone.

In the second part of this paper, I show that Nut's roles in the afterlife and autumn bird migration are consistent with similar roles attributed to the Milky Way by different cultures across the world. Moreover, echoes of Nut and her role in the Solar cycle may still be found in contemporary conceptions of the Milky Way throughout Africa. While my cross-cultural arguments do not provide any sort of proof that the ancient Egyptians linked Nut to the Milky Way, they are consistent with a growing body of work that shows how, time and again throughout history, different peoples around the world conceptualized the Milky Way in startlingly similar ways. Connecting Nut to the Milky Way would situate ancient Egypt comfortably within this framework.

## 6 ACKNOWLEDGEMENTS

I thank Juan Antonio Belmonte, Noga Ganany, Sherin Sadeq el-Gendi, Amanda-Alice Maravelia, the Pawnee Nation Tribal Historic Preservation Office, Ivo Seitenzahl, and the anonymous referees for helpful discussions and comments. I also thank the Bibliothèque nationale de France, the Los Angeles County Museum of Art (LACMA), and the Metropolitan Museum of Art for use of images from their collections, and The Egypt Exploration Society for permission to publish figure 2. This research has made use of NASA's Astrophysics Data System, Internet Archive, the JSesh open-source hieroglyphics editor (Rosmorduc 2014), and the Stellarium and Cartes du Ciel planetarium programs. Supporting research data are available on reasonable request from the corresponding author. For the purpose of open access, the author has applied a Creative Commons Attribution (CC BY) license to any Author Accepted Manuscript version arising.